\documentclass{eas}
\usepackage{graphicx}
\TitreGlobal{Setting a new standard in the analysis of binary stars}
\begin{document}

\title{The BinaMIcS project: binarity and magnetism} 
\author{C. Neiner}\address{LESIA, UMR 8109 du CNRS, Observatoire de Paris, UPMC, Univ. Paris Diderot, 5 place Jules Janssen, 92195 Meudon Cedex, France}
\author{E. Alecian}\address{Institut de Plan\'etologie et d'Astrophysique de Grenoble (IPAG), UJF-Grenoble 1/CNRS-INSU, UMR 5274 du CNRS, 38041 Grenoble, France}\sameaddress{,1}
\author{the BinaMIcS collaboration}
\begin{abstract}
We present the BinaMIcS project, its goals and the first BinaMIcS
spectropolarimetric observations obtained with Narval at TBL and ESPaDOnS at
CFHT: dedicated time-series for targeted hot and cool close double-line
spectroscopic (SB2) binaries and a survey of hot close SB2 binaries. The very
first results are also presented. In particular, our first survey observations
seem to show a lack of magnetic fields in hot close binaries compared to single
stars.
\end{abstract}
\maketitle
\section{The BinaMIcS project}

BinaMIcS (Binarity and Magnetic Interactions in various classes of Stars) is an
international project led by France (PI E. Alecian), which includes over 90
observers, modellers and theoreticians from 13 countries (see 
http://lesia.obspm.fr/BinaMIcS). The goal of the BinaMIcS project is to exploit
binarity to yield new constraints on the physical processes at work in hot and
cool magnetic stars.  In particular, BinaMIcS aims at studying (1) the role of
magnetism during stellar formation, (2) magnetospheric star-star (and
star-planet) interactions, (3) the impact of tidal flows on fossil and dynamo
magnetic fields, and (4) the impact of magnetism on mass and angular momentum
transfer.

Studying binaries rather than single magnetic stars allows to better constrain
the fundamental parameters of the targets. Moreover, binaries are an ideal
laboratory to study physical processes related to star-star interactions, such
as tidal deformation, wind-wind collisions, magnetospheric coupling, tidally
excited pulsations,... Finally, they are important to understand the origin of
magnetic fields. For example, in cool stars, they allow to test the
synchronisation of the binary versus the dynamo interaction; in
intermediate-mass stars, magnetic fields are known to be anomalously rare in
binaries and this needs to be understood; and in massive stars, some theories
propose that a field could be generated in stellar mergers.

However, spectropolarimetric studies of binary stars are particularly
challenging. First, one needs to obtain spectropolarimetric measurements over
the rotation periods of both stars and over the orbital period. Second, one
needs to disentangle the intensity spectra of both components as well as the
Zeeman signatures in the polarised light to be able to characterize the magnetic
field of both stars.

\section{The BinaMIcS observing program}

The BinaMIcS project has been allocated two Large Programs (LP) of observations:
one LP of 604 hours for 4 years (February 2013 to January 2017) with  ESPaDOnS
at CFHT (PIs E. Alecian and G. Wade), and one LP of 128 hours for 2 years (March
2013 to February 2015) with Narval at TBL (PI C. Neiner). This Narval LP is
renewable for another 2 years.

BinaMIcS observes 3 samples of targets:

(1) The cool Targeted Component (cTC): this sample consists of selected cool
magnetic SB2 dwarfs, RS CVn, BY Dra, W UMa and pre-main sequence stars, with
various fundamental and orbital parameters. These targets will be observed with
a great coverage to perform a detailed characterisation of their magnetic field
as well as detailed modelling. The results will be compared with those obtained
for single stars. We will also study the variations and correlation with
eccentricity.

(2) The hot Targeted Component (hTC): this sample consists of all 13 known 
magnetic SB2 systems of O, B or A spectral type visible from CFHT or TBL. This
includes 1, 6 and 6 systems for which the hottest star is an O, B and A star,
respectively. These targets will be observed with a great coverage to perform a
detailed characterisation of their magnetic field as well as detailed modelling.
The results will be compared with those obtained for single stars. We will also
study the wind and magnetospheric interactions between the 2 components. 

(3) The hot Survey Component (SC): this sample consists of over 200 close SB2
binaries, including eclipsing ones, with at least one component of O, B, A or
early F spectral type and a secondary component not later than F. The goal is to
search for a magnetic field in these targets. Those systems that will be
discovered to be magnetic will be transfered to the hTC sample. We will derive
the statistical occurence of magnetic fields in this sample and compare it with
the results obtained by the MiMeS collaboration for single stars
(\cite{wade2013}). 

\section{First targeted observations}

\begin{figure}[!ht]
\begin{center}
\resizebox{6.9cm}{!}{\includegraphics[clip]{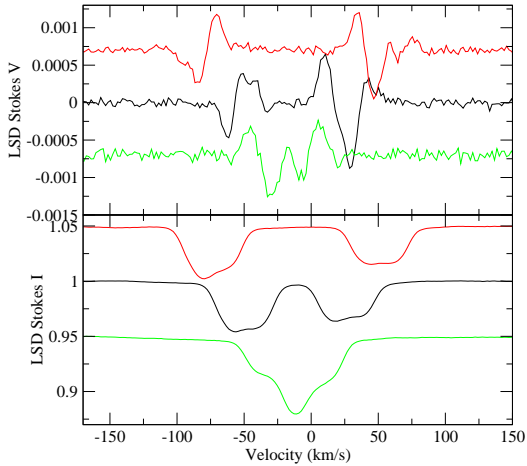}} 
\end{center}
\caption{Examples of 3 spectropolarimetric observations of the cool system
$\sigma^2$ CrB. Top: LSD Stokes V profiles showing the magnetic signatures in
both stars. Bottom: Corresponding LSD intensity profiles.}
\label{sig2crb} 
\end{figure} 

The BinaMIcS observations started in February 2013. A  complete dataset has
already been obtained for several TC stars. For the cTC, we observed BH CVn, a
RS CVn (F2IV+K2IV) system for which only one of the 2 components is magnetic,
and BY Dra (K6Ve+...) for which both components are found to be magnetic.
Snapshots have also been obtained for a number of other cool systems (UZ Tau E,
$\sigma^2$ CrB, V1379 Aql, ER Vul, HD\,216489, HD\,28, HD\,34029,...) to allow
the selection of the best cTC targets. $\sigma^2$ CrB has been selected for
follow-up as a cTC target since the first spectropolarimetric observations show
clear magnetic detections in both components (see Fig.~\ref{sig2crb}). For the
hTC, we already observed the Plaskett star, a system discovered to be magnetic
by the MiMeS collaboration (Grunhut et al., these proceedings), as well as
HD\,5550, an Ap+A system.  These TC datasets are currently beeing analysed and a
full characterisation of the magnetic fields of each system will be published in
2014. 

\section{Preliminary survey results}

\begin{figure}[!ht]
\begin{center}
\resizebox{8.8cm}{!}{\includegraphics[clip]{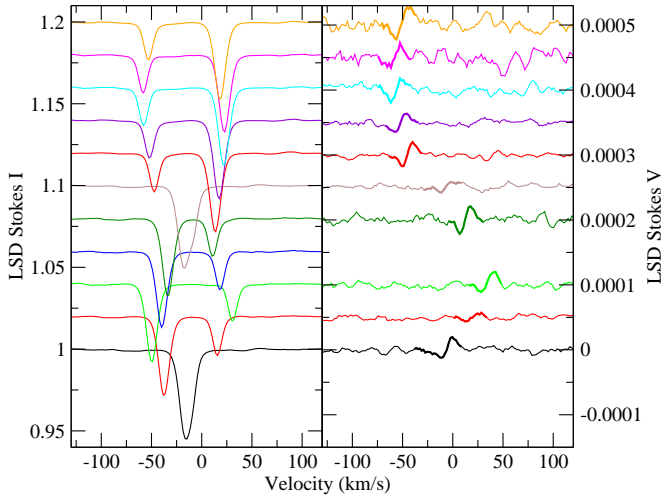}} 
\end{center}
\caption{Examples of spectropolarimetric observations of the F4+F5 system
HD\,160922. Left: LSD intensity profiles. Right: Corresponding LSD Stokes V
profiles showing the magnetic signatures in one star.}
\label{hd160922} 
\end{figure} 

At the time of writing, 93 close SB2 systems including at least one O, B, A or
early F star have been observed. Among these systems, a magnetic field has been
detected in one system: HD\,160922. This is a F4+F5 system with an orbital
period of 5.28 days. Only one of the two stars in the system shows a magnetic
signature. This target has been transferred to the hTC sample and we have
acquired 15 measurements. The simple magnetic signature does not seem to vary
much from one observation to another. Therefore this star could host a fossil
field rather than a dynamo field (see Fig.~\ref{hd160922}).

The detection of only one magnetic system among the 93 systems observed so far
raises many questions. If we consider only the 79 systems including at least one
O, B or A star, we can compare the occurence of a magnetic field in these
systems with the occurence established by the MiMeS collaboration for single
massive stars. In single stars, the occurence is 7\% (\cite{wade2013}). If the
occurence was the same in binaries, we should have detected between 5 and 11
magnetic systems. Instead we found none. Although the statistics of our current
sample is not sufficient to draw firm conclusions, the lack of detections in
this sample may be an important result for stellar formation and the origin of
magnetism in massive stars. The observations of the complete SC sample will
provide firmer conclusions and cast new light on these issues.

\section{Conclusions}

BinaMIcS is an ambitious project to study magnetic interactions in hot and cool
close binary systems, understand the role of magnetism during stellar formation
and the origin of magnetic field in massive stars. The first BinaMIcS
observations already provide very interesting results.\\

\noindent {\bf Acknowledgements}

Based on observations obtained at the Telescope Bernard Lyot (TBL)  operated by
the Observatoire Midi-Pyr\'en\'ees, Universit\'e de Toulouse, Centre National de
la Recherche Scientifique (CNRS) of France, and at the Canada-France-Hawaii
Telescope (CFHT)  operated by the National Research Council of Canada, the
Institut National des Sciences de l'Univers of CNRS, and the University of
Hawaii.


\end{document}